# Direct observation of topological surface state in the topological superconductor 2$M$-WS$_2$

Soohyun Cho[@,*,†,‡] Soonsang Huh[@,¶] Yuqiang Fang[@,†] Chenqiang Hua,[§] Hua Bai,[§] Zhicheng Jiang,[†] Zhengtai Liu,[†,‡] Jishan Liu,[†,‡] Zhenhua Chen,[∥] Yuto Fukushima,[¶] Ayumi Harasawa,[¶] Kaishu Kawaguchi,[¶] Shik Shin,[¶] Takeshi Kondo,[¶,⊥] Yunhao Lu,[*,§] Gang Mu,[*,†] Fuqiang Huang,[*,#] and Dawei Shen[*,†,‡]

[†]*State Key Laboratory of Functional Materials for Informatics, Shanghai Institute of Microsystem and Information Technology, Chinese Academy of Sciences, Shanghai 200050, China*

[‡]*Center of Materials Science and Optoelectronics Engineering, University of Chinese Academy of Sciences, Beijing 100049, China*

[¶]*Institute for Solid State Physics, University of Tokyo, Kashiwa 277-8581, Japan*

[§]*Zhejiang Province Key Laboratory of Quantum Technology and Device, Department of Physics, Zhejiang University, Hangzhou 310027, China*

[∥]*Shanghai Synchrotron Facility, Shanghai Advanced Research Institute, Chinese Academy of Sciences, Shanghai 201204, China*

[⊥]*Trans-scale Quantum Science Institute, University of Tokyo, Tokyo 113-0033, Japan*

[#]*State Key Laboratory of High Performance Ceramics and Superfine Microstructure, Shanghai Institute of Ceramics Chinese Academy of Science, Shanghai, 200050, People's Republic of China*

[@]*State Key Laboratory of Rare Earth Materials Chemistry and Applications, College of Chemistry and Molecular Engineering, Peking University, Beijing, 100871, People's Republic of China*

E-mail: shcho@mail.sim.ac.cn; luyh@zju.edu.cn; mugang@mail.sim.ac.cn;huanfq@mail.sic.ac.cn; dwshen@mail.sim.ac.cn

## Abstract

The quantum spin Hall (QSH) effect has attracted extensive research interest because of the potential applications in spintronics and quantum computing, which is attributable to two

i

conducting edge channels with opposite spin polarization and the quantized electronic conduc- tance of $2e^2/h$. Recently, $2M$-WS$_2$, a new stable phase of transition metal dichalcogenides with a $2M$ structure showing an identical layer configuration to that of the monolayer 1T′ TMDs, was suggested to be a QSH insulator as well as a superconductor with critical transition temper- ature around 8 K. Here, high-resolution angle-resolved photoemission spectroscopy (ARPES) and spin-resolved ARPES are applied to investigate the electronic and spin structure of the topological surface states (TSS) in the superconducting $2M$-WS$_2$. The TSS exhibits charac- teristic spin-momentum-locking behavior, suggesting the existence of long-sought nontrivial $Z_2$ topological states therein. We expect that $2M$-WS$_2$ with co-existing superconductivity and TSS might host the promising Majorana bound states.

**KEYWORDS**: topological surface state, ARPES, spin-resolved ARPES, 2M-WS$_2$, monolayer1T′ TMDs, superconductor

Layered transition metal dichalcogenides (TMDs) of group VIB, MX$_2$ (M = Mo, W; X = S, Se, Te), which host variant intriguing quantum states of matter such as Weyl semimetal,[1–5] topo- logical insulator,[6–10] quantum spin Hall (QSH) effect ,[11–14] valley Hall effect,[15,16] and supercon- ductivity,[17–20] have aroused great interest in both fundamental and applied research. Moreover, the family of MX$_2$ material have been discovered in a variety of stable phases with different stacking configurations (such as 3R, 2H 1T, 1T′ and 1T$_d$), which provide a unique knot to fine-tune properties of MX$_2$ by simply regulating the number of layers and stacking configurations. At present, few-layer or even monolayer MX$_2$ with a relatively large area (~ hundreds of microns) has been readily prepared through reliable means such as the mechanical exfoliation,[9,18,21] chemical vapordeposition,[22,23] and molecular beam epitaxy.[6–8]

Among the family of MX$_2$ materials, monolayer 1T′ structured TMDs (MoTe$_2$ and WTe$_2$) were suggested as promising candidates to realize QSH in strictly two-dimensional (2D) systems ,[11–14] in contrast to those previously reported quantum-well heterostructures based on three-dimensional (3D) semiconductors such as HgTe/CdTe[24–27] and InAs/GaSb.[28,29] Such QSH insulators are expected to harbor two conducting edge channels with quantized electric conductance $2e^2/h$ and opposite spin polarizations, which are robustly protected by the time- reversal symmetry.[11–14,24–29] In terms of the electronic structure, the quantized electric conductance is attributed to the topologically protected edge modes, which are generated by the strong spin-orbit coupling (SOC) induced band inversion. In this regard, topological edge states are hallmarks of QSH.[30] However, it is difficult to distinguish between electronic states from the bulk and surface for thin flakes,[31] and the characteristic features of edge states in monolayer 1T′-$MX_2$ have not yet been fully investigated experimentally in photoemission measurements.

Recently, a new meta-stable $2M$-$WS_2$, in which '$M$' indicates the monoclinic crystallographic structure, has been synthesized, hosting both enchanting topologically surface state and supercon- ductivity.[17,32–41] It is noteworthy that $2M$-$WS_2$ shares the identical monolayer crystal structure with that of 1T′-$WS_2$, while its stacking has a different packing to the 1T′ phase, as illustrated in Fig. 1(a). In other words, the bulk $2M$-$WS_2$ is regarded as the simple stacking of infinite 1T′-$WS_2$ monolayers, which are weakly coupled by the van der Waals interaction. Thus, the band structure obtained from surface sensitive photoemission spectroscopy on $2M$-$WS_2$ might shed light on the characteristic $Z_2$ topological invariant and inverted band gap in monolayered 1T′-$WS_2$.[17,32,33,35,36] In addition, $2M$-$WS_2$ behaves as a type-II superconductor with the record-high superconducting transition temperature ($T_c$) of 8.8 K among all reported TMDs,[17,34,38–41] and its extremely large magnetoresistance[35] and Majorana bound states have been reported recently.[37] Therefore, this material was considered as a promising candidate for achieving the topological superconductivity and QSH insulator due to the coexistence of superconductivity and topological non-trivial surface states.

Herein, we employ both the high-resolution and spin-resolved angle-resolved photoemission spectroscopy (ARPES) to investigate the electronic structure and helical spin texture of topological surface states (TSSs) of $2M$-$WS_2$, respectively. Our photoemission spectra show a clear $k_z$ dependence of band dispersion which can demonstrate the surface localized state, and our model was derived using a hybrid approach that combines density-functional theory calculations and symmetry considerations. Without SOC, the bulk bands with different parities overlap, forming the Dirac band crossing along the Γ-$Y$ direction. While the strong SOC would lift the degeneracy at the Dirac point, leading to an inverted band gap due to the preservation of the inversion and time- reversal symmetries. Thus, it can allow to realize the existence of topological surface states in the inverted band gap of $2M$-$WS_2$. The surface state exhibits no $k_z$ dependence inside the inverted bulk band gap, which is consistent with that predicted by the band calculation. Additionally, the spin polarization of $2M$-$WS_2$ is elucidated to show the spin helical texture in the TSS. The spin polarization along the $y$-direction $P_y$ has an opposite sign between −$k$ and +$k$ momentum points, but $P_x$ and $P_z$ are negligible. Our results show the helical spin texture of the TSS in the topological superconductor 2M-WS$_2$.

2M-WS$_2$ crystallizes in a layered monoclinic structure with the $C_{2/m}$ space group (No. 12), in which the inversion symmetry is conserved. The corresponding lattice constants are $a$ = 12.87 Å, $b$ = 3.23 Å, $c$ = 5.71 Å, $\alpha = \gamma = 90°$ and $\beta = 112.9°$, respectively.[17] Figure 1(a) displays the unit cell of the monoclinic crystallographic structure of 2M-WS$_2$ along the $b - c$ (top view) and $a - c$ (side view) planes; zig-zag chains of W atoms along the $b$-direction and layer stacking configuration. The unit cell of 2M-WS$_2$ consists of two S-W-S sandwich monolayers, which are identical to those of 1T′-WS$_2$ but are stacked along the out-of-plane *a* direction. In contrast, the manner of stacking for 1T′-WS$_2$ involves a 180° rotation about *c*/2 and *c*/2 translation along the *c*-direction as shown in Fig. 1(a). In this way, it draws a close analogy between bulk 3*R* phase and monolayer 2*H* phase MX$_2$.[42] The crystal structure of 3*R* phase shares the same monolayer of 2*H*-MX$_2$ but stacks such layers along the out-of-plane direction with a pure translation operation, so that bulk 3*R*-MX$_2$ should retain a similar broken inversion symmetry to that of monolayered 2*H*-MX$_2$. Moreover, bulk 3*R*-MX$_2$ exhibits spin band splitting for the opposite sign at different valleys, which is unique and can thus demonstrate the valley Hall effect in monolayer 2*H*-MX$_2$.[42]

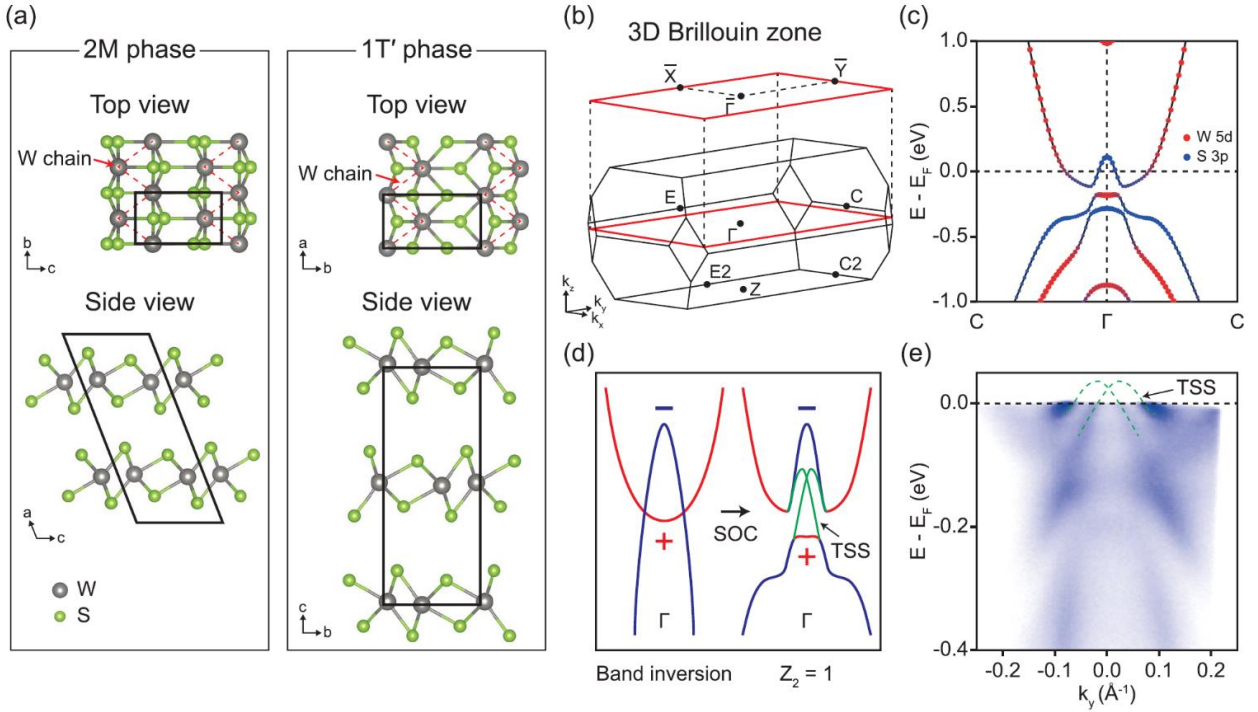

Figure 1: (a) Ball and stick representation of the 2M (monoclinic) and 1T′ (orthorhombic) structure of TMDs: a top and side view. For both crystallographic phase of monoclinic and orthorhombic symmetry, the red dotted line indicates the W-W zig-zag chain and black line is representative for the unit cell in top and side view. (b) Projected 2D surface and 3D bulk Brillouin zone with high symmetry point of the monoclinic structure. (c) The electronic structure with orbital character and with SOC along the Γ-C direction. Red (blue) circle denotes the 5$d$ (3$p$) orbital of W (S) atoms. (d) Schematic picture of band structure for bulk 2M phase and monolayer 1T′ phase of TMDs without and with SOC at the Γ point. The red (blue) line represents the + (-) parity and the green line denotes the possible shape of the TSS. (e) The ARPES intensity of 2M-WS$_2$ measured with 6.994 eV laser. The band dispersion from the $\bar{\Gamma}$ to $\bar{Y}$ direction shows the lower Dirac cone branch of the hole-like TSS.

Figure 1(b) shows the 3D bulk Brillouin zone (BZ) of monoclinic 2*M*-WS$_2$. The surface BZ is projected to the $k_x$-$k_y$ plane, and $\bar{\Gamma}$ -$\bar{Y}$ ($k_y$) corresponds to the W-W chain direction (*b*-axis). Figure 1(c) displays the calculated band structure of 2*M*-WS$_2$ with the orbital characteristics along the *C*-Γ-*C* high symmetry direction, akin to that of the monolayered 1T′-MX$_2$, one well-known QSH insulator in TMDs.[6,11–14,43] Our calculations imply that the conduction band minimum (valence band maximum) of 2*M*-WS$_2$ is mainly attributed to the *d* (*p*) orbital of W (S) atoms, with an even (odd) parity around the Γ point, respectively. With SOC, considering the symmetry of 2*M*-WS$_2$, bands with different parity will be mixed, opening the inverted band gap around the zone center.

Figure 1(d) further illustrates schematically such SOC-induced topological phase transition. Given the distortion from the 1T to 1T′ phase, the band inversion occurs around the Γ point, and band crossings give rise to Dirac cones along the $k_y$-axis. With the SOC considered, the degeneracy at these crossing points will be lifted, which results in the gap opening between the inverted bands. The symmetry and band inversion, together with strong SOC, allow the topological phase transition to a Z$_2$ insulator, which guarantees the topologically protected edge states. It is expected that these edge states will connect the electron and hole pockets, which is similar behavior with predicted surface state in the monolayer 1T′-MX$_2$.[6–10,43] Figure 1(e) shows the photoemission intensity plot of 2*M*-WS$_2$ along the $k_y$-direction. Through comparing with the calculation, it was found that the Dirac point of the topological non-trivial surface band should be above the Fermi level and the hole-like feature is the lower part of the Dirac cone. Thus, our finding confirms the existence of the TSS of 2*M*-WS$_2$ and such a surface band is markedly reminiscent of the predicted edge mode in monolayered 1T′-MX$_2$.[6–14]

Figures 2 (a) and 2 (b) compare our experimental Fermi surface mappings taken for $k_z = 0$

($h\upsilon$ = 33 eV) and $\pi$ ($h\upsilon$ = 45 eV) planes with corresponding calculations, respectively. For the $k_z$ = 0 plane (Fig. 2(a)), two electron pockets intersect with the Fermi level, which resembles the Fermi surface topology of monolayered 1T′-MX$_2$.[6,8–10,43] The electron pocket is found to be elongated in the $\Gamma$-$E$ direction, compared with that of the monolayered 1T′-MX$_2$, and consists of only electron pocket with a bulk state at the Fermi surface. Given the $k_z$ dependence of states near the Fermi level, the Fermi surface taken in the $k_z = Z$ plane shows hole pockets along both $Z$-$E2$ and $Z$-$C2$ directions as marked in Fig. 2(b). For both $k_z$ = 0 and $\pi$ cases, there exists evident spectral weight around the zone center which is absent in bulk band calculations. Furthermore, regardless of $k_z$ values, such a feature in the bulk band gap show the similar behavior as shown in the $k_x$-$k_y$ plane. Both pieces of evidence suggest that this feature originated from the TSS of 2M-WS$_2$.

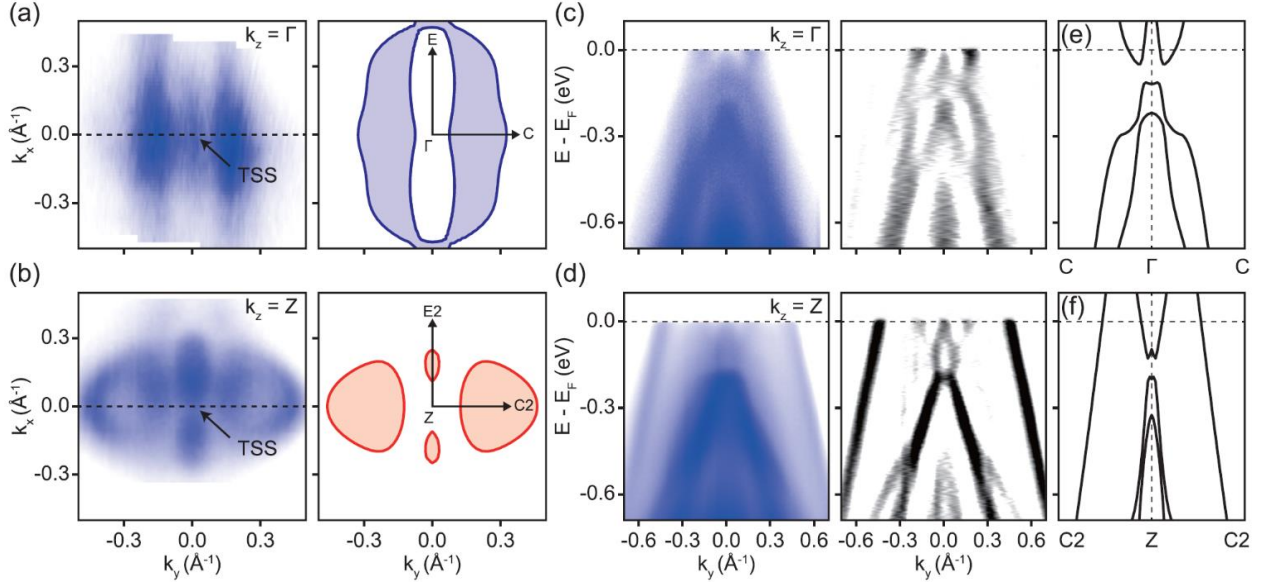

Figure 2: Fermi surface at the $k_z = \Gamma$ (a) and the $Z$-plane (b). Left (right) panel can be obtained from the ARPES measurement (bulk band calculation). The $k_x$ ($k_y$) refers to the $\bar{\Gamma}$ -$\bar{X}$ ($\bar{\Gamma}$ -$\bar{Y}$) direction. The blue (red) line at the $\Gamma$ ($Z$) plane indicates the electron (hole) pocket. The ARPES spectra at $k_z = \Gamma$ (c) and $Z$ point (d) along the black dotted line from (a) and (b). The right-hand panel in (c) and (d) show second derivative spectra of the corresponding left-hand panels to enhance the visualization. The bulk band calculation at the $\Gamma$ (e) and $Z$-plane (f).

More detailed low-energy band dispersions taken along cuts marked by dashed black lines in Figs. 2(a) and 2(b) are displayed in Figs. 2(c) and 2(d). Moreover, the corresponding second derivative intensity plots are demonstrated in the right-hand panels of Figs. 2(c) and 2(d). Both band dispersions exhibit good agreement with the calculations of the bulk bands as shown in Figs. 2(e) and 2(f). For $k_z = 0$ (Fig. 2(c)), there exist extra features around the zone center in the vicinity of the Fermi level besides the predicted bulk electron-like bands; while for the $k_z = \pi$ plane, the bulk bands crossing the Fermi level become hole-like. Given the strong SOC in 2$M$-WS$_2$, the inverted band gap is located between the electron pocket as the CBM and hole pocket as the VBM near the $\bar{\Gamma}$ point. This band gap is induced by the band inversion and lifting the state degeneracy along the $k_y$-direction, resulting in the emergence of the TSS inside the bulk band gap near the Fermi level. Besides, we performed the detailed polarization-dependent ARPES measurements on this compounds, as illustrated in Supporting Information. We note the orbital character of 2$M$-WS$_2$ along the $\Gamma$-$Y$ direction is almost identical to that of the monolayer 1T′-WTe$_2$,[6,43] which further implies that the monolayer 1T′-WS$_2$ would be a promising candidate for the quantum spin Hall insulator.

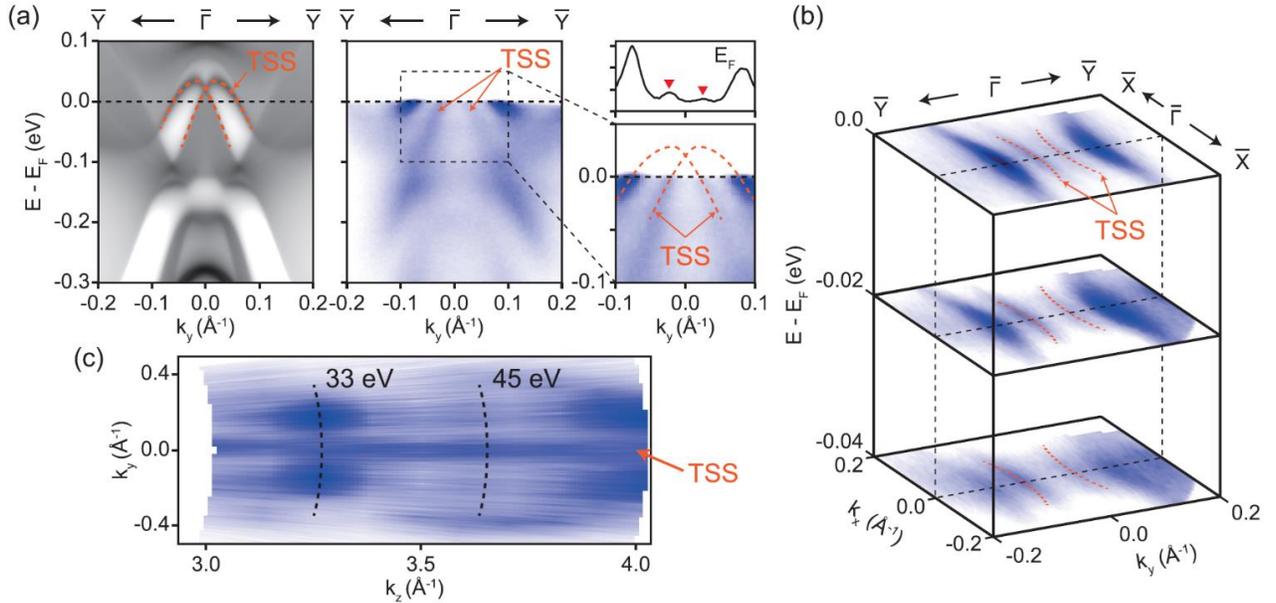

Figure 3: Electronic structure of the non-trivial TSS in 2$M$-WS$_2$. (a) The band structure obtained from the slab calculation (left) and laser based ARPES measurement (middle) along the $\bar{Y}$ -$\bar{\Gamma}$ -$\bar{Y}$ direction. The TSS highlighted by the orange dotted line show the hole-like dispersion and the Dirac points of the TSS exist above the Fermi level. Zoomed in dispersion of the TSS is from the black box area in the middle panel of (a). The momentum distribution curves were taken from the zoomed in dispersion of the TSS at the Fermi level. The red reversed triangles indicate the $k_F$ of the TSS. The distance between the $k_F$s of the TSS is approximately 0.045 Å$^{-1}$. (b) The staking plots of the ARPES constant energy map at the different binding energy show the evolution of the TSS. (c) The Fermi surface of the $k_z$-$k_y$ plane, which covers more than one periodicity of BZ size. The 33 and 45 eV photon energy are corresponding to the $k_z = \Gamma$ and $Z$ planes, respectively.

Next, the slab calculations along G-Y were calculated to demonstrate the details of the TSS around the zone center (Fig. 3(a)). Given the existence of the inverted gap around $\bar{\Gamma}$ close to the Fermi level, the distinct feature in the gap is attributed to topologically protected helical surface band, which is also consistent with previous reports.[17,32,33] This topological surface band connects the VBM with the 3$p$ orbital (− parity) and the CBM with the 5$d$ orbital (+ parity) so that it shows the hole-like dispersion as marked with orange lines in Fig. 3(a). It should be noted that these calculations, particularly for the surface bands, are in good agreement with our experimental band structure shown in Fig. 3(a). Figure 3(b) shows the evolution of the hole-like dispersion of the TSS in the constant energy contours at the different binding energy.

Then, detailed photon-energy dependent ARPES measurements ($27 \leq h\upsilon \leq 54$ eV) on 2$M$-WS$_2$ were performed to investigate the topological surface band (for details, see Supporting Information). It is noted that the probing photon energy range can cover more than one BZ along $k_z$. As shown in Fig. 3(c), the clear periodic pattern is observed in the $k_z$-$k_y$ intensity map. Given this periodicity, $k_z \approx 1.06$ Å$^{-1}$ ($k_z = 2\pi/a^*$, where $a^*$ is 5.93 Å) and an inner potential of 12.5 eV can be obtained. $a^*$ is determined by the perpendicular distance between the layer of 1$M$-WS$_2$ as given by $a^* = (a/2) \sin(\beta)$. The length of $a^*$ attributable to the periodicity of $k_z$ and inner potential is determined by the shortest distance between the 1$M$-WS$_2$ layers, not the length of a in the mono-clinic unit cell of 2$M$-WS$_2$. In contrast to the $k_z$ dispersion of the bulk band, the TSS has typical quasi-two-dimensional band dispersion along the $k_z$-direction and the spectral weight remain at $k_y = 0$ over more than one periodicity of BZ because the surface state is localized in the surface of the 2$M$-WS$_2$.

The topological non-trivial surface band should be spin-polarized, as suggested by our first-principles calculations. Afterwards, spin-resolved ARPES was used to realize this prediction. The low-energy photoemission spectra obtained from the laser-based ARPES with 6.994 eV photons clearly displays the fine dispersion of the TSS in $k_x$-$k_y$ plane with the higher momentum resolution in Fig. 4(a). The TSS dispersion at Fermi surface is good agreement with that obtained from the slab calculation in Fig. 4(b). The hole-like band dispersion near $k_y = 0$ corresponds to the lower part of the Dirac cone of TSS, and the band crossings are expected to be located above the Fermi level. To characterize the helical spin structure of the TSS, we utilize the spin-resolved ARPES equipped with the double very-low-energy electron diffraction spin detectors for the 3D spin-resolution along the $x$, $y$, and $z$-axis, and the photon-electron intensity attributable to the different magnetization directions $M_+$ and $M_-$ for the $x$, $y$, and $z$-directions can be obtained. With the relationship (given the details of the experimental method), the spin-resolved energy distribution curves (EDCs) $s_x$, $s_y$, and $s_z$, were extracted respectively. Figure 4(c) shows the spin-resolved EDCs of $s_y$ with red and blue colors assigned for spin-up and spin-down, respectively, which indicate that the sign of $s_y$ is reversed between $k_L$ and $k_R$.

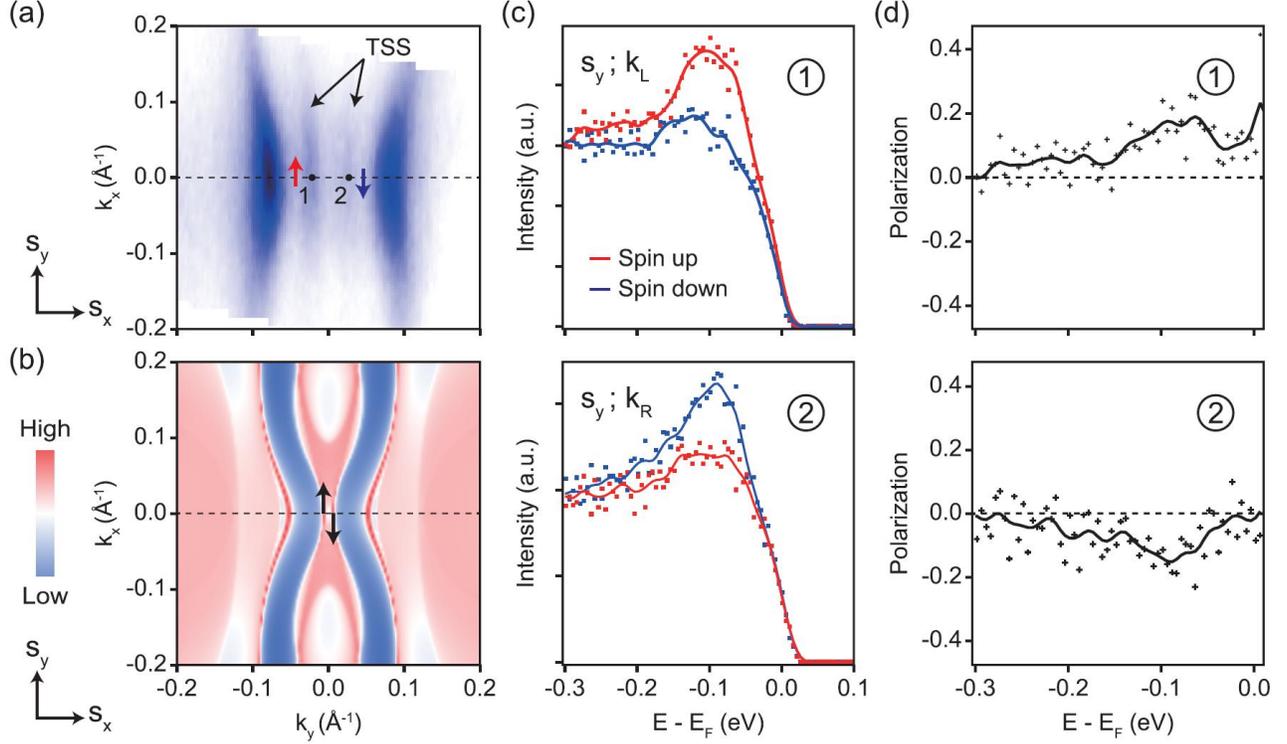

Figure 4: The spin-momentum-locked TSS taken from laser-based SARPES with 6.994 eV photons. (a) Fermi surface with the in-plane spin direction in $k_x$-$k_y$ plane. The red (blue) arrow indicates the spin up (down) along the $s_y$ direction. (b) The Fermi surface obtained from the photoemission measurement have a good agreement with the calculated Fermi surface in (b). The black arrow indicate the in-plane spin direction of the TSS based on the band calculation. (c) Spin-resolved EDCs of the TSS at the left and right momentum points as marked in (a). The spin-resolved EDCs at $k_L$ and $k_R$ represent spin texture along the $s_y$ direction. (d) Representative spin-polarization magnitudes for the TSS at $k_L$ and $k_R$, which were obtained from the spin-resolved EDCs in (c).

To reveal the difference of spin momentum at $k_L$ and $k_R$, the magnitude of spin-polarization $P_y$ attributable to the spin-resolved EDCs $s_y$ was determined as shown in Fig. 4(d). The in-plane spin-polarization $P_y$ is reversed between $k_L$ and $k_R$ near the Fermi level in Fig. 4(d), but another spin-polarization $P_x$ and $P_z$ are negligible (the spin-polarization map of $P_x$, $P_y$ and $P_z$ can be seen in Supporting Information). It is expected that the TSS of 2$M$-WS$_2$ has the helical spin texture, thus allowing the opposite spin direction between left and right momenta at the Fermi level. Note that our experiments about the spin-polarization successfully show the spin-momentum-locked TSS in 2$M$-WS$_2$ as illustrated in Fig. 4(d). In the view of the Fermi surface based on our experimental results, the lower Dirac cone branch of the TSS in 2$M$-WS$_2$ has a helical spin texture with a clockwise rotation (for details of calculation, see Supporting Information). Our spin-resolved ARPES results are good agreement with the usual topological feature[44] and the band calculation in Fig. 4(b). Given the crystal and electronic structural similarity between 2$M$-WS$_2$ and monolayer 1T′-MX$_2$,[6,8–10,17,33,43] it is expected that the spin-resolved ARPES results of the TSS in 2$M$-WS$_2$ provide important information about the edge conductance and spin-based device as manufactured with QSH insulator, monolayer 1T′-MX$_2$ and about the topology of the 2$M$ phase of TMDs.

To conclude, both the high-resolution and spin-resolved ARPES were used to investigate the electronic structure and helical spin texture of TSSs of 2$M$-WS$_2$, respectively. The results indicate that the TSS is located in the inverted band gap, consistent with the calculation. Additionally, the spin polarization along the y-direction $P_y$ has opposite signs between $k_L$ and $k_R$ momentum points, but $P_x$ and $P_z$ are negligible. Our results show the helical spin texture of the TSS in the topological superconductor 2$M$-WS$_2$. Thus, the band structure obtained from surface-sensitive photoemission spectroscopy on 2$M$-WS$_2$ may shed light on the characteristic Z$_2$ topological invariant and inverted band gap in monolayered 1T′-WS$_2$. Besides, this material can be considered as a promising candidate for realizing the topological superconductivity and QSH insulator due to the coexistence of superconductivity and topological non-trivial surface states.

## Method

ARPES measurements were performed at both 03U beamline of Shanghai Synchrotron Radiation Facility (SSRF).[45] The end-station was equipped with Scienta-Omicron DA30 electron analyzer. The angular and the energy resolutions were set to 0.2° and 8 ~ 20 meV (dependent on the selected probing photon energy). All samples were cleaved in an ultra-high-vacuum better than $8.0 \times 10^{-11}$ torr. All ARPES data were taken from the horizontal polarized light as measured at 10 K.

The laser based spin-resolved ARPES (SARPES) measurements were carried out at the Institute for Solid State Physics (ISSP), University of Tokyo, which was equipped with 6.994 eV laser.[46] All spectra were acquired with a 6.994 eV laser a ScientaOmicron DA30-L analyzer, to gether with double very-low-energy electron diffraction (VLEED) spin detectors. The energy resolution for the SARPES measurements was set to about 20 meV. The SARPES measurements were taken from horizontal polarized light at 10 K. To obtain the spin-resolved EDCs from the SARPES measurements, we use the following equation:[46]

$$s_\uparrow^\alpha = \frac{1}{2}(1 + P^\alpha)(M_+^\alpha + M_-^\alpha)$$

$$s_\downarrow^\alpha = \frac{1}{2}(1 - P^\alpha)(M_+^\alpha + M_-^\alpha)$$

where $\alpha$ denotes the resolved axis ($x$, $y$, and $z$-axes), and $M^\alpha$ ($M^\alpha$) represents the photon-electron intensity obtained from the SAPRES measurements with different directions of magnetization + and − (up and down) for the double VLEED spin detectors along the $\alpha$-axis. $P^\alpha$ is representative of the spin polarization along the $\alpha$-axis and can be obtained by:

$$P^\alpha = \frac{1}{S_{eff}}\frac{M_+^\alpha + M_-^\alpha}{M_+^\alpha + M_-^\alpha}$$

where $S_{eff}$ is an *effective Sherman function*, and it is 0.28 in our spin-resolved ARPES mea-surements.[46] The maximum magnitude of spin-polarization $P^\alpha$ is ±1.

Our calculations were performed using the Vienna ab initio simulation package (VASP)[47] with the projector-augmented wave (PAW) method.[48] The generalized gradient approximation with Perdew, Burke, and Ernzerhof (PBE)[49] realization was used for the exchange-correlation functional. The energy cut-off was set above 400 eV and the force and energy convergence criteria were set to 0.001 eV/Å and $10^{-8}$ eV, respectively. A 10 × 10 × 7 Γ-centered $k$-point mesh was used for the Brillouin zone sampling. The van der Waals corrections were included by the dispersion- corrected density functional theory calculations.[50] The surface states were calculated by the Wannier90 package[51] and the WannierTools code.[52] The SOC was considered and the experiment lattice constants were taken in all calculations.[17]

## Author Contributions

@ Soohyun Cho, Soonsang Huh and Yuqiang Fang contributed equally in the work.

## Notes

The authors declare no conflict of interest.

## Acknowledgement

This work was supported by National Science Foundation of China (Grant Nos. U2032208 and 11974307), National Key R&D Program of China (2019YFE0112000), Zhejiang Provincial Natural Science Foundation of China (LR21A040001), and Natural Science Foundation of Shanghai (Grant No. 14ZR1447600). J.S.L. thanks the fund of Science and Technology on Surface Physics and Chemistry Laboratory (6142A02200102). Part of this research used Beamline 03U of the Shanghai Synchrotron Radiation Facility, which is supported by ME$^2$ project under Contract No. 11227902 from National Natural Science Foundation of China.

## Supporting Information Available

The Supporting Information is available free of charge at

- Filename: Spin-up and spin-down spectra along the $S_x$, $S_y$ and $S_z$ of the topological surface state, polarization-dependent ARPES study, detailed photon-energy-dependent study of the ARPES spectra, the calculated spin texture of the topological surface state at the Fermi surface.

# Supporting Information for
# "Direct observation of topological surface state in the topological superconductor 2$M$-WS$_2$"

Soohyun Cho,[1,2,*] Soonsang Huh,[3,†] Yuqiang Fang,[4,5,†] Chenqiang Hua,[6] Hua Bai,[6] Zhicheng Jiang,[1] Zhengtai Liu,[1,2] Jishan Liu,[1,2] Zhenhua Chen,[7] Yuto Fukushima,[3] Ayumi Harasawa,[3] Kaishu Kawaguchi,[3] Shik Shin,[3] Takeshi Kondo,[3,8] Yunhao Lu,[6,‡] Gang Mu,[1,§] Fuqiang Huang,[4,5,¶] and Dawei Shen[1,2,**]

[1]*State Key Laboratory of Functional Materials for Informatics,*
*Shanghai Institute of Microsystem and Information Technology,*
*Chinese Academy of Sciences, Shanghai 200050, People's Republic of China*
[2]*Center of Materials Science and Optoelectronics Engineering,*
*University of Chinese Academy of Sciences, Beijing 100049, People's Republic of China*
[3]*Institute for Solid State Physics, The University of Tokyo, Kashiwa, Chiba 277-8581, Japan*
[4]*State Key Laboratory of High Performance Ceramics and Superfine Microstructure,*
*Shanghai Institute of Ceramics Chinese Academy of Science, Shanghai, 200050, People's Republic of China*
[5]*State Key Laboratory of Rare Earth Materials Chemistry and Applications,*
*College of Chemistry and Molecular Engineering, Peking University, Beijing, 100871, People's Republic of China*
[6]*Zhejiang Province Key Laboratory of Quantum Technology and Device,*
*School of Physics, Zhejiang University, Hangzhou 310027, People's Republic of China*
[7]*Shanghai Synchrotron Facility, Shanghai Advanced Research Institute,*
*Chinese Academy of Sciences, Shanghai 201203, People's Republic of China*
[8]*Trans-scale Quantum Science Institute, The University of Tokyo, Bunkyo-ku, Tokyo 113-0033, Japan*

PACS numbers:

# Contents

# 1. Spin-up and spin-down spectra of 2$M$-WS$_2$

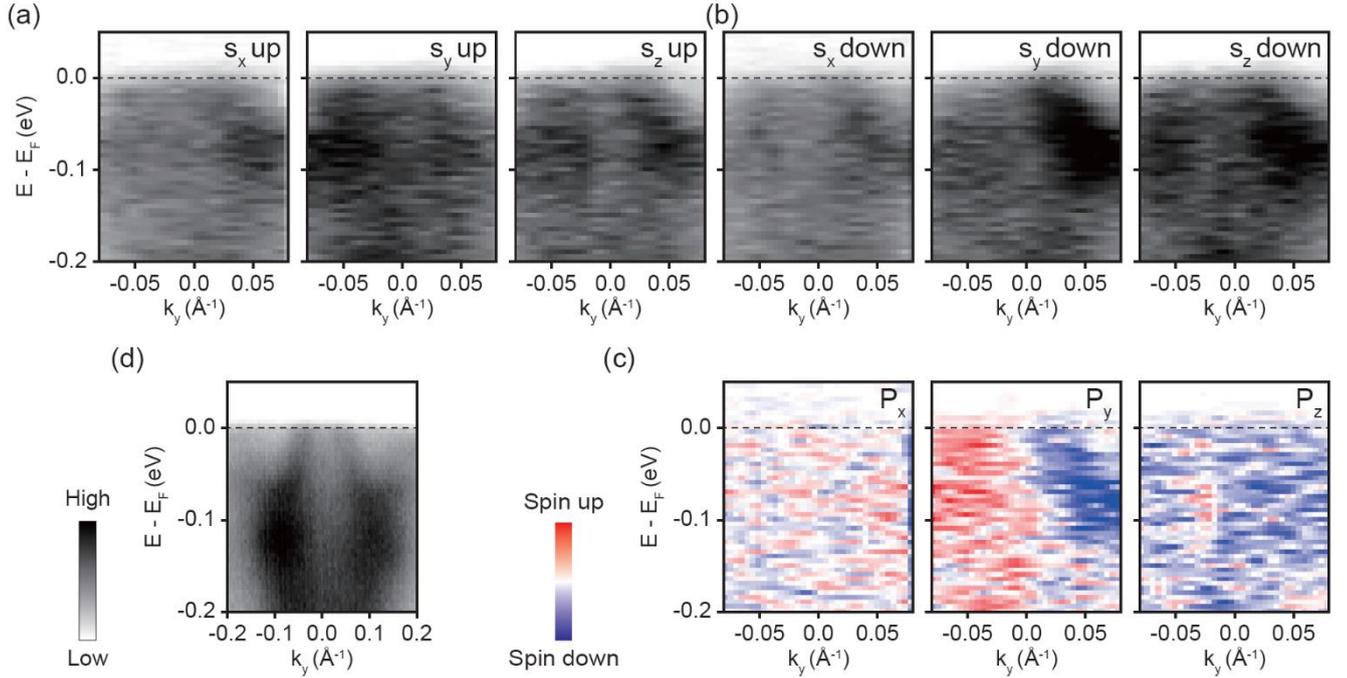

**FIG. S1:** The spin-resolved ARPES spectra of topological surface state in 2$M$-WS$_2$. (a) Spin-up and (b) spin-down spectra in the $s_x$, $s_y$ and $s_z$ direction. (c) Spin-polarization in the $P_x$, $P_y$ and $P_z$ direction, which obtained from the spin-resolved ARPES spectra of the TSS in (a) and (b). The red (blue) color indicate the spin-up (-down). While the spin-polarization in $P_y$ have opposite sign along the $k = 0$, the spin-polarization in $P_x$ and $P_z$ are negligible. (d) The ARPES intensity of 2$M$-WS$_2$ is taken from 6.994 eV laser along the $\bar{Y} - \bar{\Gamma} - \bar{Y}$ direction. The electronic structure shows the lower Dirac cone branch of the topological surface state near the $\Gamma$-point.

## 2. Polarization dependent ARPES measurements.

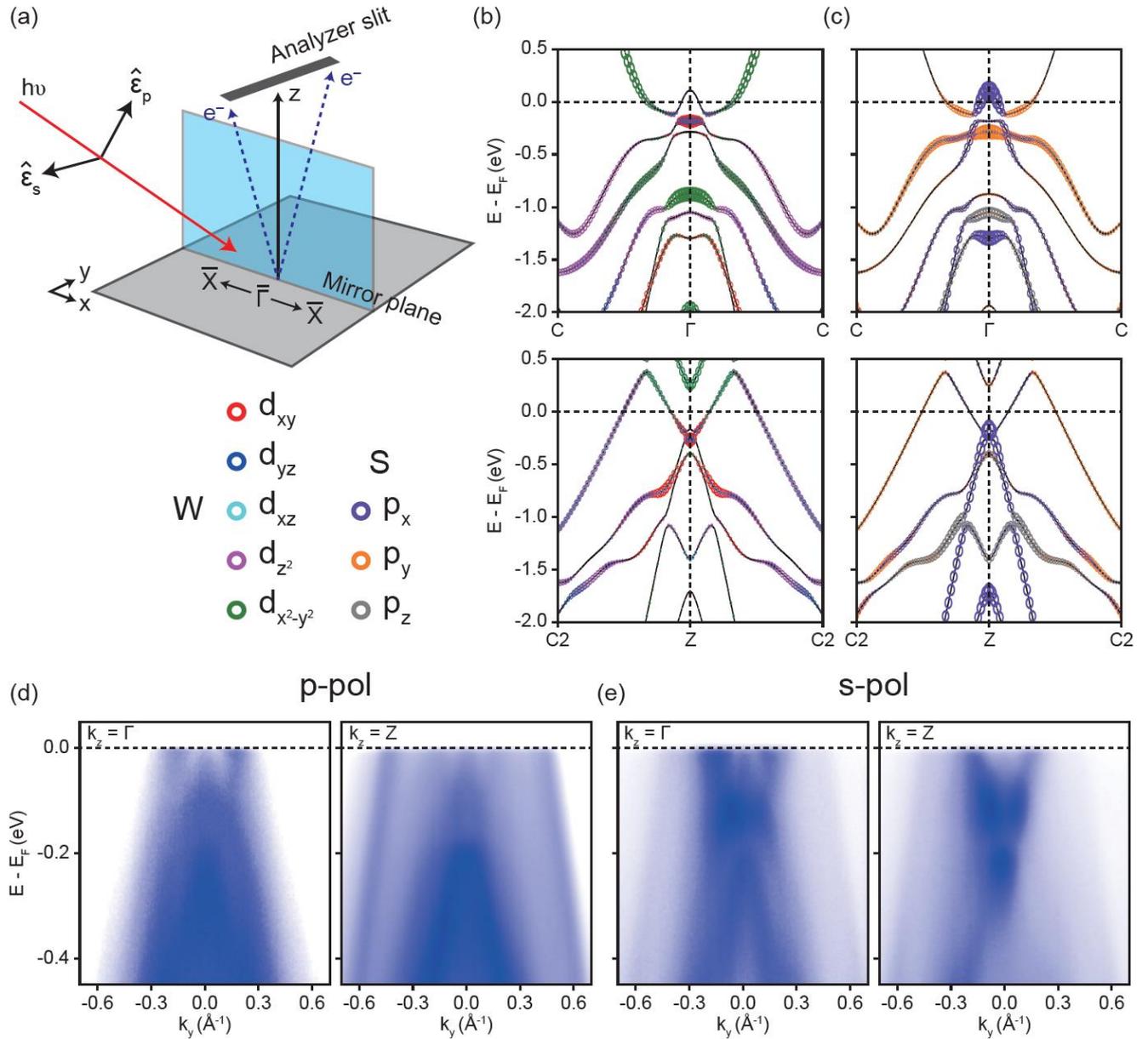

**FIG. S2:** Polarization dependent ARPES measurements. (a) Experimental geometry with the mirror plane including the incident light. The analyzer slit is vertical to the mirror plane and parallel to the Γ-$Y$ direction. The $\hat{\varepsilon}_s$ and $\hat{\varepsilon}_p$ are a representative of the unit vector with the s- and p-polarized light, respectively. (b, c) The bulk band calculation at the $k_z$ = Γ and Z plane is along the Γ-$Y$ direction. The calculated valence and conduction band from $d$-orbitals of W atoms (b) and $p$-orbitals of S atoms (c) in 2M-WS$_2$. The bulk band near the Fermi level is manly contributed by the $d_{z^2}$, $d_{xy}$ and $d_{yz}$ orbital of W atoms and the $p_x$ and $p_y$ orbital of S atoms. The ARPES spectra at the $k_z$ = Γ (33 eV) and Z (45 eV) plane along the $\bar{\Gamma} - \bar{Y}$ direction, which were taken with the p-polarized (d) and s-polarized (e) light.

|  | parity | orbital character | | | | |
|---|---|---|---|---|---|---|
| s-polarization | odd | $d_{xy}$ | $d_{yz}$ | $p_y$ | | |
| p-polarization | even | $d_{xz}$ | $d_{x^2-y^2}$ | $d_{z^2}$ | $p_x$ | $p_z$ |

**TABLE I:** Relationship of the parity between the polarization dependent photoemission spectroscopy and the orbital character.

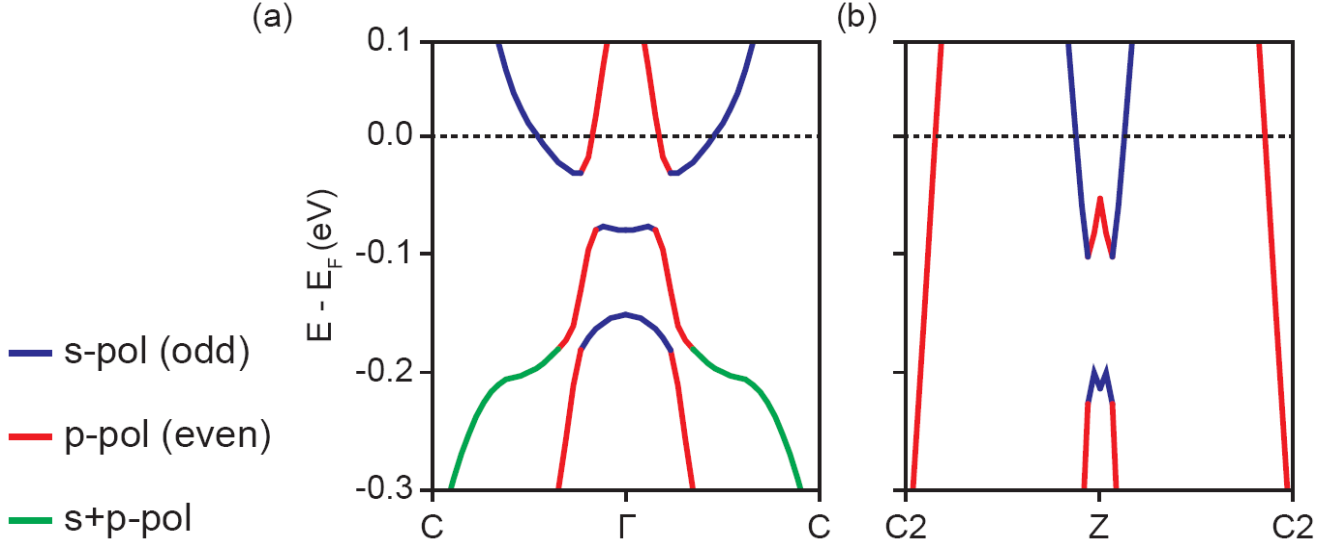

**FIG. S3:** Schematic picture of the orbital selection. The band dispersion along the $C$-$\Gamma$-$C$ (a) and $C2$-$Z$-$C2$ (b) direction. The color coded with blue (red) indicate the orbital character with e odd (even) parity and the higher residual ARPES intensity taken with s- (p-) polarized light, respectively.

We can briefly show the relationship of the orbital parity between the ARPES measurement with the linear polarization and the orbital character in Table 1. The even- (odd-) parity of the initial states can be observable with the p- (s-) polarization following that the squared one-electron matrix elements in photoemission spectroscopy are proportional to $|M_{f,i}^k|^2 \propto |\langle \phi_f^k | \hat{\varepsilon} \cdot \hat{x} | \phi_i^k \rangle|^2$, where the $\hat{\varepsilon}$ is a unit vector along the polarization direction of the vector potential **A** and $\hat{\varepsilon}_s$ ($\hat{\varepsilon}_p$) are a representative of the unit vector with the s- (p-) polarized light, respectively. Given the matrix elements, the non-vanishing photoelectron intensity can be determined by the relationship between the orbital parity of the initial-state crystal wave function and the polarization direction of the vector potential. Given that the final state $\varphi^k$ itself was even, an even function for the overlap integral between the initial state and the polarization direction allows the non-vanishing photoemission intensity.

Figure S2(d) and (e) show that the ARPES spectra taken with the s- (p-) polarized light have a relatively high residual spectral weight composed with the orbital character of $d$ and $p$ orbitals with the odd (even) parity with respect to the mirror plane, respectively. The valence band at the $k_z = \Gamma$ and $Z$ plane are manly contributed by the $d_{z^2}$, $d_{x^2-y^2}$, $d_{xy}$, and $d_{yz}$ orbital of W atoms, and the $p_x$ and $p_y$ orbital of S atoms in $2M$-WS$_2$. The ARPES spectra taken with the p-polarized light in Fig. S2(d) show the higher ARPES intensity of the bulk state contributed by the $d_{z^2}$ and $d_{x^2-y^2}$ orbital of W atoms and $p_x$ orbital of S atoms, which is an even parity with respect with the mirror plane. The bulk state with the odd parity for the $d_{xy}$, $d_{yz}$, $p_y$ orbitals have a relatively high residual spectral weight in the s-polarized light $\hat{\varepsilon}_s$ in Fig. S2(e). The ARPES spectra taken with s-polarized light in Fig. S2(e) is similar to the band structure for ARPES spectra obtained from the synchrotron-based ARPES measurement in previous report [1]. For the briefly summary, Figure S3 show the schematic picture for the bulk state at the $k_z = \Gamma$ and $Z$ plane, together with the parity of the orbital character and the polarization direction. The color coded with the blue (red) in the calculated bulk band indicate the non-vanishing photoemission intensity determined by the initial state and s- (p-) polarized light.

## 3. Photon energy dependent ARPES measurement.

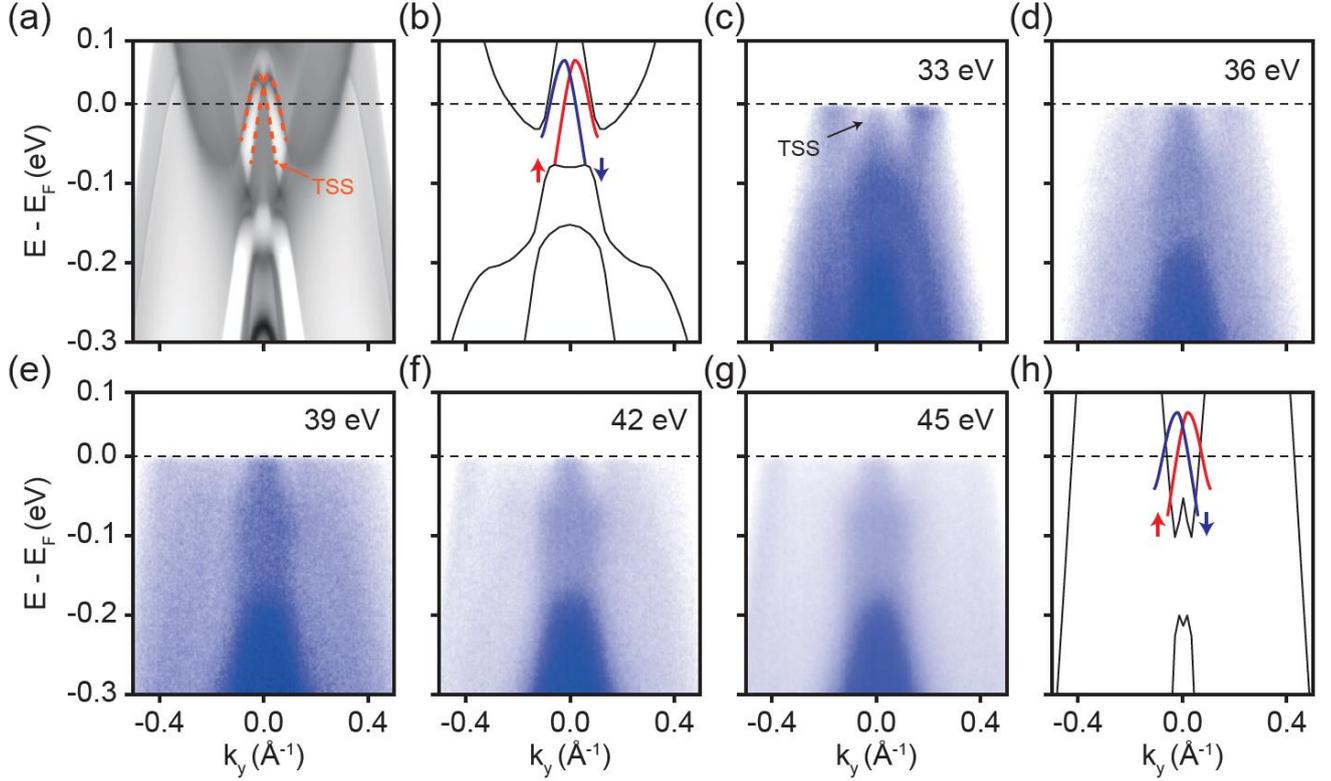

**FIG. S4:** Photon energy dependent ARPES spectra. (a) The surface projected calculations of 2$M$-WS$_2$. The orange line indicates the TSS. The schematic picture for the electronic structure with TSS at the $k_z$ = Γ-plane (b) and Z-plane (h). The blue and red lines indicate the possible TSS with in-plane $s_y$ spin up and down directions, respectively. (c)-(g) The ARPES spectra along the $\bar{\Gamma} - \bar{Y}$ direction at the photon energy from 33 to 45 eV.

Figures S4 (b)-(h) show the evolution of electronic structure with the photon energy from 33 to 45 eV, corresponding to the Γ and Z-planes along the $k_z$-direction, respectively. The bulk state near the Fermi level indicates strong $k_z$ dependence, but the TSS remains intact regardless of the photon energy. The TSS at 36 and 39 eV seems to show a disconnected state from the bulk state, indicating that the state exists in the bulk band gap. This feature of the TSS is consistent with the band calculation with the surface state in Fig. S4(a).

## 4. Band calculation for the spin direction of the TSS.

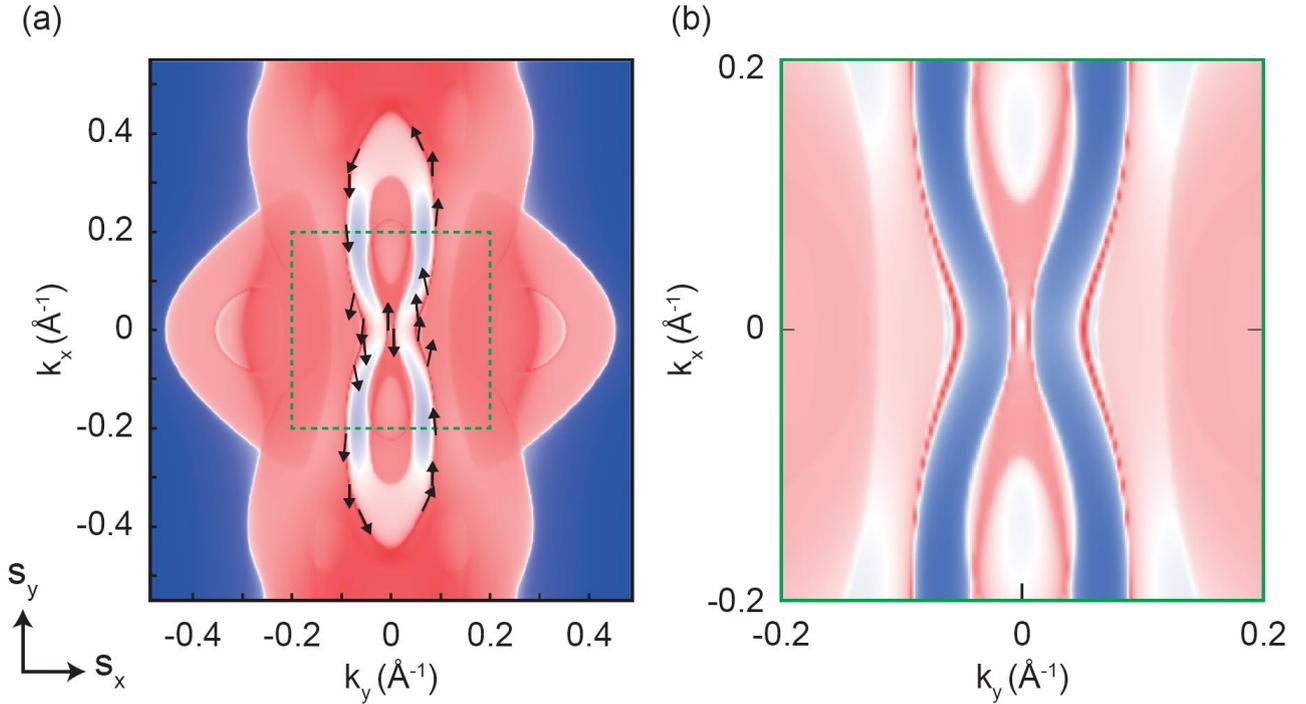

**FIG. S5:** Calculation for the spin-momentum locking of the TSS on the Fermi surface. (a) The Fermi surface obtained from the band calculation show the spin direction of the TSS on the $k_x$-$k_y$ plane. The black arrow indicates the in-plane spin direction. The spin direction of $s_x$ and $s_y$ about the lower Dirac cone of the TSS along the $\Gamma$-$Y$ direction have the spin momentum approximately the order of $\pm 10^{-6}$ and $\pm 0.533$ (normalization to $\pm 1$) at the $k_F$, respectively. It indicates that the in-plane spin momentum is fully aligned along the $s_y$ direction. (b) Enlarged picture for the TSS within the green dotted box in (a).

## References

∗ Equal contributions; Electronic address: shcho@mail.sim.ac.cn

† Equal contributions

‡ Electronic address: luyh@zju.edu.cn

§ Electronic address: mugang@mail.sim.ac.cn

¶ Electronic address: huanfq@mail.sic.ac.cn

∗∗ Electronic address: dwshen@mail.sim.ac.cn